\documentclass[a4paper]{spie}  
\usepackage{subcaption}
\usepackage{float}
\usepackage{amsmath,amsfonts,amssymb}
\usepackage{graphicx}
\usepackage[colorlinks=true, allcolors=blue]{hyperref}

\newcommand{\apj}{\textit{The Astrophysical Journal}}
\newcommand{\aap}{\textit{Astronomy \& Astrophysics (A\&A)}}
\newcommand{\solphys}{\textit{Solar Physics}}

\title{In orbit background for hard X-ray CubeSat polarimeters: case study of the CUSP mission in low-earth orbit}

\author[a]{Abhay~Kumar}
\author[a,c]{Giovanni~Lombardi}
\author[g,n]{Riccardo~Campana}
\author[g]{Giovanni~De~Cesare}
\author[a]{Sergio~Fabiani}
\author[a]{Ettore~Del~Monte}
\author[a,b]{Andrea~Alimenti}
\author[a]{Enrico~Costa}
\author[h]{Mauro~Centrone}
\author[a]{Nicolas~De~Angelis}
\author[a]{Sergio~Di~Cosimo}
\author[a]{Giuseppe~Di~Persio}
\author[a]{Pasqualino~Loffredo}
\author[l]{Gabriele~Minervini}
\author[a]{Fabio~Muleri}
\author[m]{Paolo~Romano}
\author[a]{Alda~Rubini}
\author[a]{Emanuele~Scalise}
\author[a,b]{Enrico~Silva}
\author[a]{Paolo~Soffitta}

\author[i]{Giovanni~Cucinella}
\author[i]{Vito~Di~Bari}
\author[i]{Simone~Di~Filippo}
\author[i]{Andrea~Negri}
\author[i]{Massimo~Perelli}

\author[j,k]{Dario~Modenini}
\author[j]{Andrea~Curatolo}
\author[k]{Nicolas~Gagliardi}
\author[j]{Daniele~Pecorella}
\author[k]{Alice~Ponti}
\author[j,k]{Paolo~Tortora}

\author[d]{Andrea~Del~Re}
\author[d]{Davide~Albanesi}
\author[d]{Valerio~Campamaggiore}
\author[d]{Giulia~de~Iulis}
\author[d]{Adam Ebrahim}
\author[d]{Paolo~Leonetti}
\author[d]{Marco Schepis}
\author[d]{Alessandro~Zambardi}
\author[e,o]{Ilaria~Baffo}
\author[e,o]{Pierluigi~Fanelli}
\author[e]{Costantino~Zazza}

\author[f]{Valerio~Vagelli}
\author[f]{Daniele~Brienza}
\author[f]{Immacolata~Donnarumma}
\author[f]{Matteo~Mergè}
\author[f]{Emanuele~Zaccagnino}
\author[f]{Alessandro~Turchi}

\affil[a]{INAF-IAPS, Via del Fosso del Cavaliere 100, 00133 Rome, Italy}
\affil[b]{Department of Industrial, Electronic and Mechanical Engineering, Roma Tre University, Via V. Volterra 62, 00146, Roma, Italy}
\affil[c]{Department of Enterprise Engineering ”Mario Lucenti”, University of Rome Tor Vergata, Via Cracovia 50, 00133 Rome, Italy}
\affil[d]{DEDA Connect s.r.l., Via Vincenzo Lamaro 51, 00173 Rome, Italy}
\affil[e]{DEIM, University of Tuscia, Largo dell’Università, 01100 Viterbo, Italy}
\affil[f]{ASI, Via del Politecnico snc, 00133, Roma, Italy}
\affil[g]{INAF-OAS Bologna, Via Gobetti 93/3, 40129, Bologna, Italy}
\affil[h]{INAF-OAR, Via Frascati 33, 00040, Monte Porzio Catone, Italy}
\affil[i]{IMT s.r.l., via Carlo Bartolomeo Piazza 30, 00161 Rome, Italy}
\affil[j]{Department of Industrial Engineering - Alma Mater Studiorum University of Bologna - Via Montaspro 97, 47121 Forlì, Italy}
\affil[k]{Interdepartmental Centre for Industrial Aerospace Research - Alma Mater Studiorum University of Bologna - Via Carnaccini 12, 47121 Forlì, Italy}
\affil[l]{INAF-Headquarters, Viale del Parco Mellini 84, 00136, Roma, Italy}
\affil[m]{INAF-OACT, Via S. Sofia 78, 95123, Catania, Italy}
\affil[n]{INFN Bologna Section, viale Berti Pichat 6/2, 40127, Bologna, Italy}
\affil[o]{DIBAF, University of Tuscia, Largo dell'Università, 01100, Viterbo, Italy}

\authorinfo{Further author information: (Send correspondence to Abhay Kumar)\\Abhay Kumar: E-mail: abhay.kumar@inaf.it, Telephone:  +39 06 49934511}

\begin{document} 
\maketitle

\begin{abstract}

The CUbesat Solar Polarimeter (CUSP) project aims to measure the linear polarization of solar flares in the 25 – 100~keV hard X-ray band using a Compton scattering polarimeter. CUSP is a project in the framework of the Alcor Program\footnote{\url{https://www.asi.it/en/technologies-and-engineering/micro-and-nanosatellites/alcor-program/}, consulted on 12 Jul 2025} of the Italian Space Agency aimed to develop innovative CubeSat technologies and missions. As part of CUSP’s Phase B study, initiated in December 2024 and closed on July 2nd, 2026, estimating the in orbit background to optimize the signal-to-background ratio was one of the key objectives. In low-Earth orbit, the instrument is exposed to cosmic and albedo X-ray backgrounds, charged particles, and secondary radiation from the spacecraft and atmosphere. Simulating these contributions enables optimization of detector geometry and shielding to maximize signal-to-noise performance. 

We present initial in orbit background estimates for CUSP using a Geant4-based simulator. A detailed mass model of the CUSP has been implemented to simulate background components and estimate the background count rate in the CUSP orbit.
\end{abstract}

\keywords{CUSP, X-ray polarimetry, Geant4, Solar flares, Compton polarimetry, CubeSat, In orbit background, Polar orbit, Space Weather, Heliophysics}

\section{INTRODUCTION}
\label{sec:intro}  

The Sun is our closest star and the main source of energy for life on Earth. It also releases energy through violent, high-energy explosions known as solar flares. During these flares, magnetic field lines reconnect, releasing massive amounts of energy. This accelerates particles along magnetic field lines, into interplanetary space and down into the lower solar atmosphere. These events can affect space weather and affect technology on Earth. Even after studying the Sun for over a century, two major mysteries remain. It is still not fully understood what physical process creates the Sun's magnetic field and how its outer atmosphere the corona gets so hot. The corona reaches temperatures over one million Kelvin, which is much hotter than the Sun's surface (the photosphere), which sits at  6,000 Kelvin \cite{klimchuk2006,moortel2015}. It is widely accepted that the magnetic field lines rising from the surface play a key role in heating the corona. The Sun's X-ray spectrum consists of emission lines at lower energies (below 10 keV), and at higher energies (above 10 keV), it has both thermal and non-thermal Bremsstrahlung components. Different theoretical models predict different polarization for thermal and non-thermal emissions \cite{emslie1980,zharkov2010}. However, these models cannot be discriminated based on the energy spectrum. Furthermore, the amount and direction of this X-ray polarization depend on how the particles are beamed, the shape of the magnetic field, and viewing angle \cite{Jeffrey2020}. The polarization measurement is essential to disentagle between different models. During the prompt phase of a solar flare, the X-ray spectrum is dominated by non-thermal radiation from fast-moving electrons. As the flare evolves, the surrounding plasma heats up, causing thermal radiation to take over—around in the 10 to 30 keV range. This reduces the overall polarization \cite{Nagasawa_2022,Grigis2004,dennis2005}. Therefore, it is important to study solar flares with sufficient time resolution to catch these rapid changes.

The actual polarisation measurements of solar flares have been few and statistically limited due to its difficulty in the measurement techniques. So far, most studies have only managed to find upper limits or borderline detections \cite{Tindo1970,Tindo1972a,Tindo1972b,Tramiel1984,suarez2006,boggs06}. To advance this field, instruments with high sensitivity and fast timing resolution is needed to capture rapidly changing flare events. The CUbesat Solar Polarimeter (CUSP)\cite{fabiani2025cubesat} instrument is designed to measure time resolved polarization in the 25 to 100 keV energy range. This will help us understand the mechanisms behind flare emissions, particles acceleration and associated magnetic field.

The space radiation environment poses a significant challenge for sensitive spaceborne instruments like CUSP which will operate in a Sun-Synchronous Orbit (SSO) at an altitude of $~500\text{ km}$ and an inclination of 97.4$^\circ$. When operating in Low Earth Orbit (LEO), a satellite is exposed to a complex matrix of primary and secondary radiation components. 
The intensity of these background components is modulated by Earth's geomagnetic field. The earth's magnetic field acts as a natural shield, deflecting incoming lower-energy solar and galactic charged particles through a mechanism quantified as geomagnetic cutoff rigidity. Because the magnetic field lines converge at the poles and are strongest at the equator, this shielding effect varies with latitude. Consequently, a satellite's orbital latitude determines its long-term background radiation exposure and ultimate instrument signal-to-noise ratio.

The activity presented is aimed to develop a Geant4 based simulator and analysis method to estimate the in orbit radiation background of the CUSP and optimise the instrument shielding. This study uses the Geant4 Monte Carlo simulation toolkit and background models \cite{campana2013background,cumani2019background} to track how the detector responds to individual background components. The analysis focuses on the instrument's 25–100 keV operational energy window, evaluating instrument performance within a high-latitude (97.4$^\circ$) and a low-latitude (0$^\circ$) LEO orbits. Since the satellite in an SSO continuously undergoes latitude transitions and encounters varying local magnetic shielding conditions throughout each 94.6-minute orbit, evaluating the background at the two extreme points (0$^\circ$ and 97.4$^\circ$) provides an upper and lower envelope of the in-orbit background in SSO orbit. The study quantifies the particle-induced background across different radiation sources. Ultimately, the resulting background spectra provide the essential baseline data needed to refine CUSP's hardware shielding, configure its event-tagging logic, and build post-flight data cleaning pipelines to isolate true solar polarimetry signals from the background.

\section{The CUSP: CUbesat Solar Polarimeter}

\begin{figure}[H]
  \centering
  \includegraphics[width=0.4\textwidth]{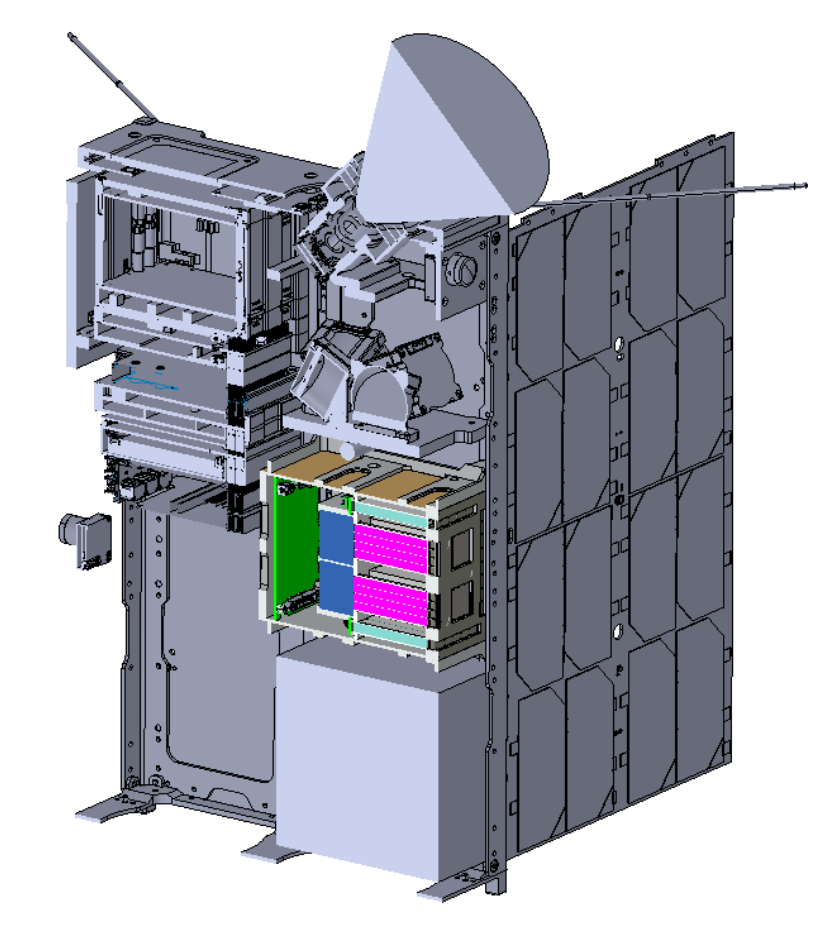} 
  \includegraphics[width=0.53\textwidth]{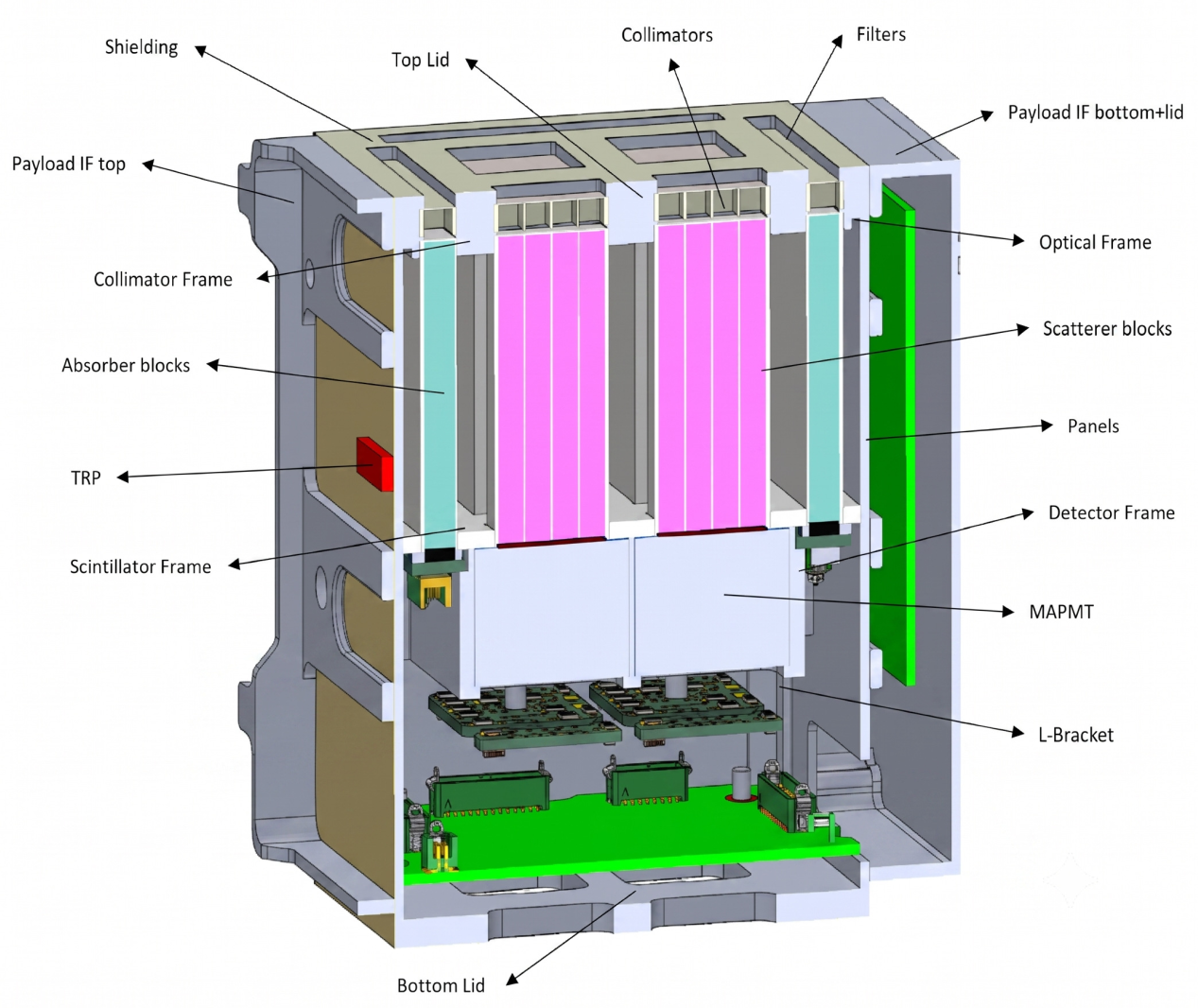}
  \caption{\textbf{Left:} The scheme of the CUSP platform with the payload at the center. \textbf{Right:} The current Phase B CAD model of the CUSP payload}
  \label{fig:cusp_design}
\end{figure}

The CUSP project, funded by the Italian Space Agency (ASI), is in the framework of the Alcor program implemented by ASI to develop innovative
CubeSats and bringing together a consortium of public and private entities, including research institutes, universities, and small to medium-sized enterprises (SMEs). The CUSP project is led by INAF-IAPS, which is responsible for the design and integration of the scientific payload in collaboration with Deda Connect s.r.l. The 6U-XL CubeSat platform is developed by IMT s.r.l., while the Interdepartmental Center for Aerospace Industrial Research (CIRI-AERO) of the University of Bologna conducts the mission analysis. The University of “La Tuscia”, Viterbo, Italy, has the responsibility of the Ground Station. Science Operations Centre (SOC) will be implemented within the ASI's Space Science Data Center (SSDC) infrastructure. Software development and data processing activities will be under INAF's responsibility, leveraging SSDC's resources and expertise for optimization and deployment. CUSP features a 6U-XL CubeSat (see Figure \ref{fig:cusp_design}) orbiting the Earth in SSO orbit (~500 km) with an inclination of approximately 97.4$^\circ$ and a period of approximately 94.6 minutes. The satellite spins at 1 RPM around the polarimeter symmetry axis pointing the Sun to allow to better control the systematic effect known as spurious modulation.

The payload comprises an X-ray polarimeter based on Compton scattering. In the CUSP energy range, Compton scattering is the dominant process of photon matter interaction. Polarised radiation induces a preferential azimuthal angular direction of scattering (normal to the incident beam axis) as described by the Klein-Nishina cross section \cite{heitler54}:

$$\frac{\mathrm{d}\sigma}{\mathrm{d}\Omega}=\frac{r_{0}^{2}}2\left(\frac{E^{\prime}}E\right)^{2}\left(\frac E E+\frac{E^{\prime}}E-2\sin^{2}\theta\cos^{2}\eta\right)$$

where,
$${\frac{E^{\prime}}{E}}=\left[1+{\frac{E}{m_{\mathrm{e}}c^{2}}}(1-\cos\theta)\right]^{-1}$$

where $r_\circ$ is the classical electron radius, m$_{e}$c${^2}$ represent the rest mass energy of an electron, E and E$^{'}$ are the incident and scattered  photon energy, respectively, $\theta$ is the Compton scattering angle, and $\eta$ denotes the angle forms between the plane of scattered photon direction with the plane containing the polarization direction of the incident photon.

X-ray photons impinging on the detector are scattered by an array of 64 plastic scintillators whose light output is detected by means of 4 Multi Anode Photomultiplier tubes (MAPMT) readout by a MAROC 3A ASIC by WEEROC. Scattered photons are finally absorbed by an array of 32 inorganic scintillators made
of GAGG:Ce ($\mathrm{Gd}_3\mathrm{Al}_2\mathrm{Ga}_3\mathrm{O}_{12}$:Ce) coupled to APD sensors readut by a SKIROC 2A ASIC by WEEROC (see Figure \ref{fig:cusp_design}). Each scintillator bar is fitted with a tungsten collimator and X-ray filter to limit its field of view to approximately $\pm$36$^{\circ}$ and effectively suppressing the background to improve detection sensitivity. Polarization of the incident X-ray photons is obtained from the cos square modulation of the azimuthal Compton scattering angle, measured in the plane perpendicular to the incident photon direction.

\section{SPACE RADIATION ENVIRONMENT}\label{bkg_models}

The Low Earth Orbit (LEO) background spectrum is highly dynamic and depends mainly on the orbit altitude and inclination, as well as the current level of solar activity. Outside the South Atlantic Anomaly (SAA), prompt background contributions are mainly: (i) primary cosmic-ray protons, alpha-particles, electrons and positrons, whose intensity is modulated by the Earth’s magnetic field, (ii) secondary protons, neutrons, electrons and positrons produced by cosmic-ray interaction with the Earth’s atmosphere, (iii) the Earth’s hard X-ray/gamma-ray albedo, and (iv) the cosmic diffuse X- and gamma-ray background. All background components are assumed to be isotropic. The background model spectral information is taken from the papers by Cumani et al. \cite{cumani2019background} and Campana et al. \cite{campana2013background}. They summarised the collected data and analyses performed by various experiments; such as Fermi-LAT, AMS-02, etc.

\begin{figure}[H]
  \centering
  \includegraphics[scale=0.5]{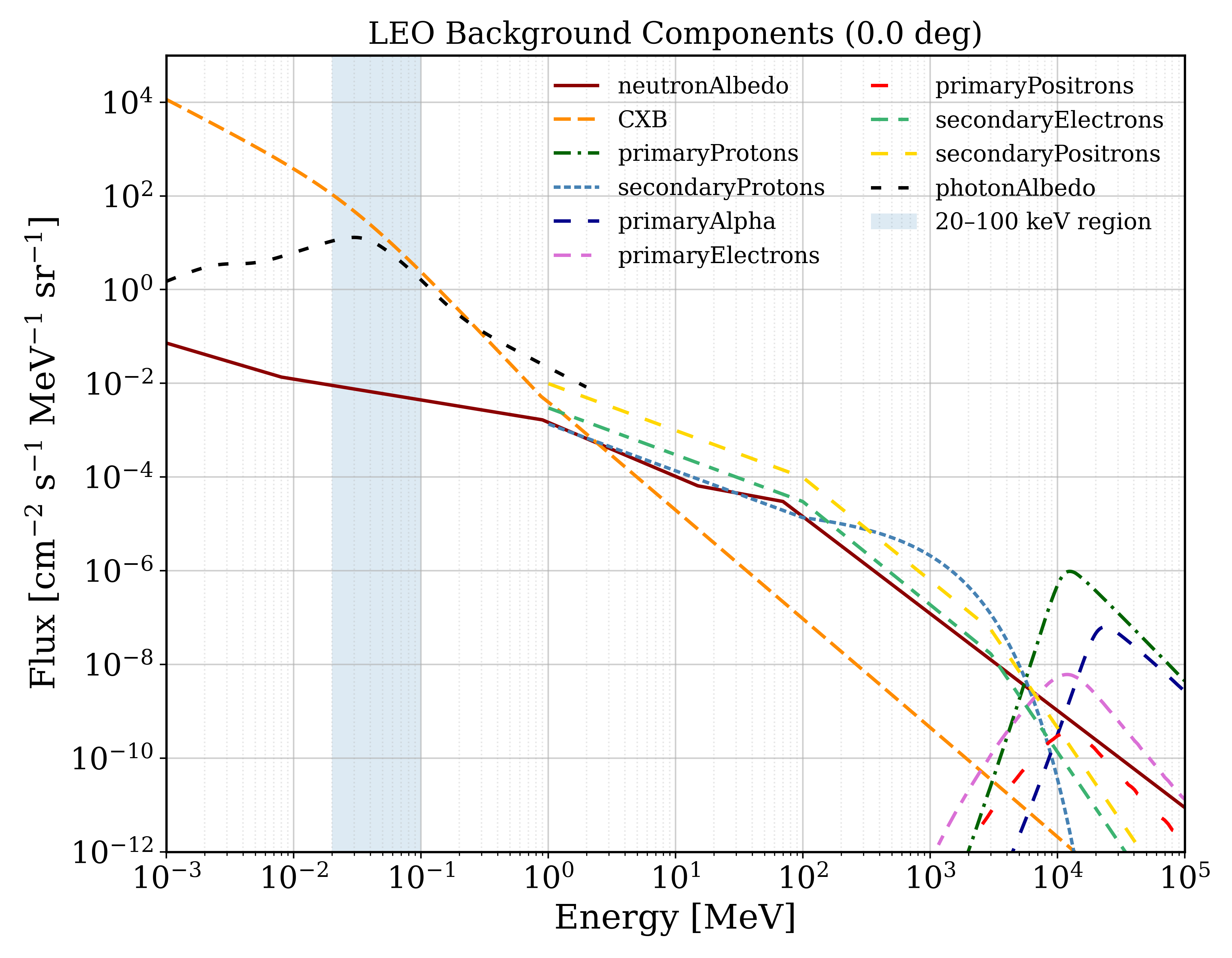} 
  \caption{The model spectra of the background components outside the SAA in LEO at 0$^\circ$ latitude.}
  \label{fig:bkg_model}
\end{figure}

The orbit altitude and inclination dictate the particle intensities. The input spectra injected through the General Particle Source (GPS) in Geant4 simulations are generated using different cutoff rigidity (Earth-shielding effects) to evaluate CUSP's performance at two latitudes (0 and 97.4$^\circ$). The geomagnetic cutoff mainly affects the lower energy particle fluxes. For 0$^\circ$ latitude, the background model (see Figure \ref{fig:bkg_model}) is dominated by the steady, isotropic flux of the diffuse Cosmic X-ray Background (CXB), as the strong geomagnetic field near the equator effectively shields the spacecraft from low-to-mid energy primary cosmic rays and their corresponding atmospheric showers \cite{campana2013background,cumani2019background}. Conversely, for 97.4$^\circ$ latitude scenario, the background model incorporates the highly elevated particle fluxes accounting for the reduction in geomagnetic cutoff rigidity at high latitudes, which allows a dense population of Galactic Cosmic Rays (GCRs) to penetrate the upper atmosphere. The model captures the resulting secondary particle environment, specifically tracking the increase of the atmospheric photon and neutron albedo fluxes. Furthermore, both models account for solar cycle modulations, as solar activity directly affects atmospheric density and alters the interplanetary magnetic field, which in turn modulates the incoming cosmic ray flux.

\section{Geant4 simulation framework}\label{sec:geant4}

Interaction of background radiation and particles with the active detector elements and surrounding passive materials of the CUSP spacecraft is simulated using a dedicated Monte Carlo simulation framework written in C++ and based on Geant4 \cite{agostinelli03_short} (version 11.2.1, installed on ubuntu 22.04 system). The geometric mass model of the satellite, the input particle spectrum, physics models describing particle interactions, general particle source and the selections on energy deposits are needed in order to determine background rates. 

To balance tracking accuracy with computational efficiency, a hybrid mass model of the spacecraft is used. The satellite platform, that lies far from the sensitive detector, is simplified into a series of squares and cubes filled with an equivalent aluminum density to accurately match the structural mass and bulk shielding properties of the spacecraft. However, the CUSP payload including detectors, collimators, and front-end electronics is modeled precisely to capture all particle interactions, shielding asymmetric designs, and true material compositions. The combined geometry was converted into the
Geometry Description Markup Language (GDML) format, allowing it to be seamlessly imported into the Geant4 simulation environment. The material properties of different GDML model components are defined in the detector construction file. A reference physics list offered by the Geant4 collaboration named the Shielding Physics List has been used in the simulation. This package includes all the physics processes needed for simulations in the space environment, which include electromagnetic physics, hadronic physics and radioactive decay physics.

\begin{figure}[H]
  \centering
  \includegraphics[scale=0.6]{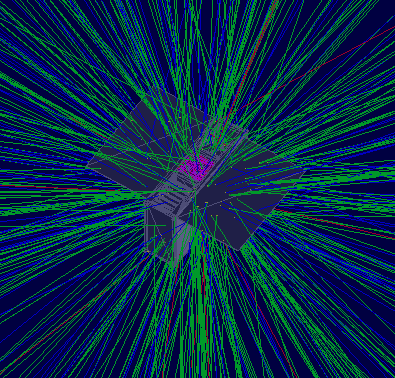} 
  \caption{CUSP mass model at the center of the GPS.}
  \label{fig:cusp_massmodel}
\end{figure}

The radiation environment is simulated by defining a General Particle Source (GPS) as a spherical surface surrounding the mass model. The combined satellite platform and payload geometry are positioned at the center of this sphere (see Figure \ref{fig:cusp_massmodel}). Particles are generated on the surface of the sphere and injected inward toward the satellite using an angular distribution that correctly models an isotropic space environment. The macros file of the primary input spectra injected through the GPS are generated based on the Cumani \cite{cumani2019background} and the Campana  \cite{campana2013background} background models as discussed in section \ref{bkg_models}.

\section{Data analysis and In orbit background}

The primary and secondary particles entering the mass model passes through the simplified aluminum equivalent satellite platform structure and active payload volume depositing energy through various mechanisms. For CUSP, only the coincidence Compton events are treated as valid polarization signals. A true Compton event is defined as an incident photon undergoes Compton scattering within the plastic scatterer and is subsequently absorbed by the GAGG absorber. Applying the same conditions to the background simulated data yields the background event. The combined total energy deposited in the Compton event is used to generate the raw simulated spectra for both orbital scenarios (inclination 0° and 97.4°)

The in-orbit background spectra are obtained by applying a Lower-Level Discriminator (LLD) of 1 keV in plastic and 24 keV in GaGG while the Upper Level Discriminator (ULD) of 16 keV in plastic and 84 keV in GaGG. Within the simulation framework, LLD and ULD are dictated by the Compton kinematics for $90^\circ$ photon scattering across the 25-100 keV target energy band of CUSP. Practically, the LLD cutoff is governed by the intrinsic low-energy detection thresholds of the plastic scatterer bars, whereas the ULD boundary is constrained by the maximum dynamic range of the readout Application-Specific Integrated Circuit (ASIC). To account for the real-world operational capabilities of the instrument's electronic triggering system, Compton tagging efficiency is applied.

\begin{figure}[H]
  \centering
  \includegraphics[scale=0.6]{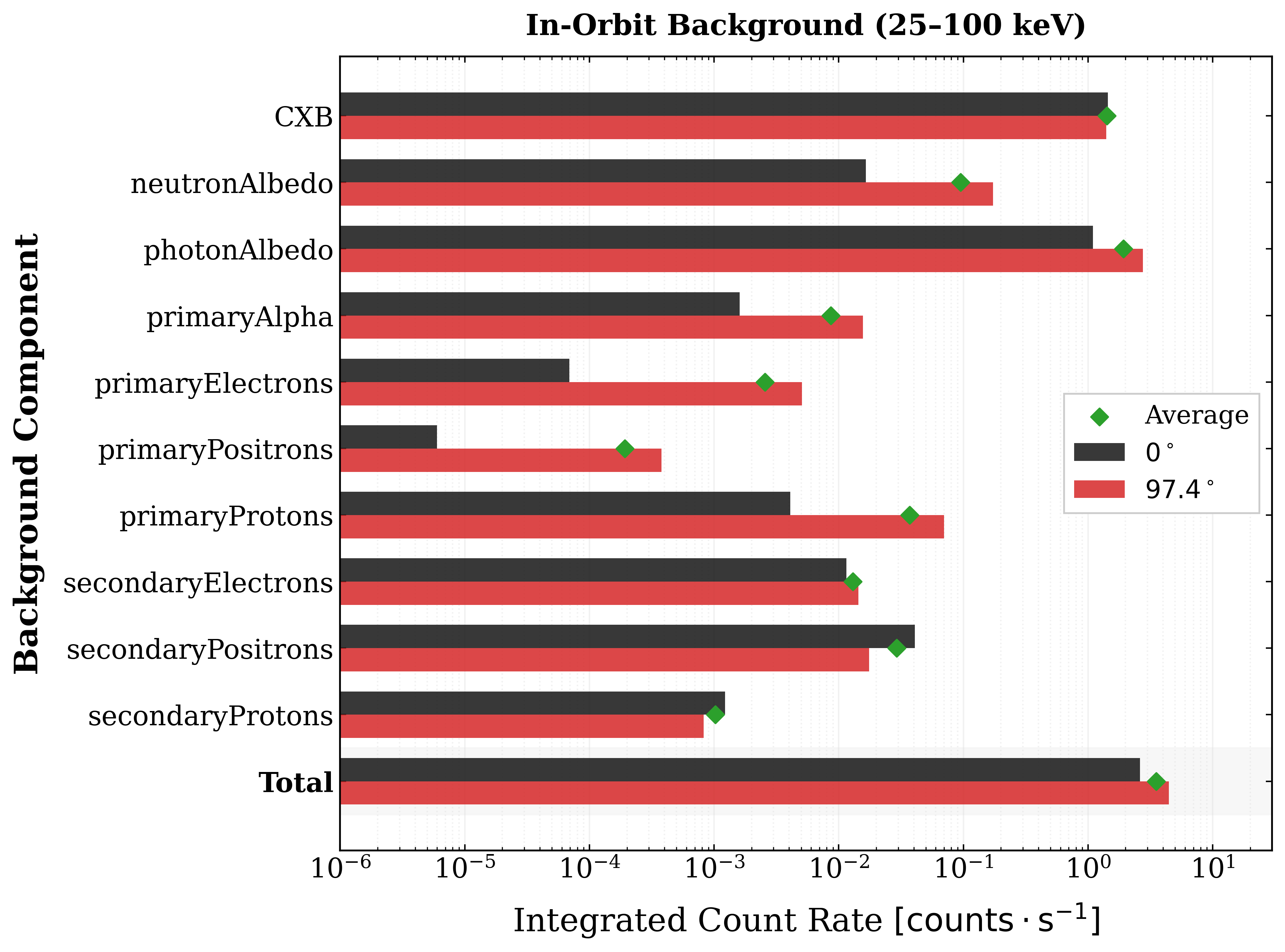} 
  \caption{Estimated background count rates
}
  \label{fig:cusp_bkg}
\end{figure}

Further, to convert the idealized Geant4 energy depositions into realistic experimental observations, the tagged energy events are blurred using an energy resolution interpolation profile measured in the lab for GAGG (Gadolinium Aluminum Gallium Garnet) scintillators. This final step produced the tagged and smeared background spectrum for each of individual background radiation components. By integrating the finalized, resolution-smeared spectra over the 25–100 keV window and multiplying them by the total geometric collection surface area. The overall background rate varies between 2.61 counts/s at the equator ($0^\circ$ latitude) and 4.44 counts/s at the poles ($97.4^\circ$ latitude). The background remains mainly dominated by CXB and photon albedo, with direct and secondary particle-induced interactions contributing just $3\%$ and $7\%$ of the total rate at $0^\circ$ and $97.4^\circ$, respectively (see Figure \ref{fig:cusp_bkg}).

\section{Conclusions and Future Work}
We developed a Monte Carlo code based on Geant4 to estimate the in orbit background of the CUSP mission using a detailed mass model of the instrument and also an analysis method to get the integrated background count rate corresponding to the different background components. This code and data analysis method will be used to optimise the shielding  of the payload as well as the signal to noise ratio for different class of Solar flares. Here, we have shown the initial estimations of in orbit background results obtained using the current mass model. As a first result, the zero degrees latitude characterized by a lower, total background count rate in 25 to 100 keV CUSP energy band. The geomagnetic cutoff near the equator suppresses primary cosmic rays, meaning the background is dominated almost entirely by the steady, faint baseline of the diffuse Cosmic X-ray Background (CXB). The low background level is in line with the expectations for an instrument based on coincident detection of signals. Such a low level is compatible with the sensitivity expectations as reported in Fabiani et al. 2026 SPIE. At latitude of 97$^\circ$, a higher total background count rate is obtained, because the magnetic shielding is weaker compared to equatorial orbit, the cosmic rays striking the atmosphere creates an intense secondary cascade (see Figure \ref{fig:cusp_bkg}).

We aim to evaluate CUSP’s scientific performance by estimating the expected Signal-to-Noise Ratio (SNR) across different classes of solar flares. In parallel, structural and shielding designs will be optimized to suppress the in-orbit background and enhance polarimetric sensitivity. To refine our radiation environment modeling, background estimations will be extended to intermediate latitudes ($\sim 50^\circ$), and activation background, specifically due to the satellite passing through the South Atlantic Anomaly (SAA) and radiation belts, will be assessed. Finally, this comprehensive background estimates will help in driving the definition of CUSP’s operational modes, establishing the safe scientific observation windows and identifying orbital segments where high-voltage (HV) bias must be reduced or switch off to ensure detector safety.

\acknowledgments 

This work is funded by the Italian Space Agency (ASI) under the Alcor program. Program, within the development of the CUbesat solar polarimeter ( CUSP ) mission under the ASI-INAF contract n. 2023-2-R.0.

\newpage
\bibliographystyle{spiebib} 

\end{document}